\documentclass[]{aastex631}
\usepackage{CJK}
\usepackage{amsmath}
\usepackage{bm}
\usepackage{float}
\usepackage{booktabs}
\usepackage{multirow}
\usepackage{epsfig}

\graphicspath{{./}{figures/}}
\usepackage{hyperref} 

\begin{document}
\begin{CJK*}{UTF8}{gbsn}

\title{Correlation between U/Th and Pb/Os abundance ratios and its application in nuclear cosmochronology}

\author{Y. Y. Huang \CJKfamily{gbsn} (黄盈予)}
\affiliation{Department of Physics, Fuzhou University, Fuzhou 350108, Fujian, China}

\author{Q. Q. Cui \CJKfamily{gbsn} (崔青青)}
\affiliation{Department of Physics, Fuzhou University, Fuzhou 350108, Fujian, China}

\author{X. H. Wu \CJKfamily{gbsn} (吴鑫辉)}
\email{wuxinhui@fzu.edu.cn}
\affiliation{Department of Physics, Fuzhou University, Fuzhou 350108, Fujian, China}
\affiliation{State Key Laboratory of Nuclear Physics and Technology, School of Physics, Peking University, Beijing 100871, China}

\author{S. Q. Zhang \CJKfamily{gbsn} (张双全)}
\affiliation{State Key Laboratory of Nuclear Physics and Technology, School of Physics, Peking University, Beijing 100871, China}

\begin{abstract}
The abundance ratios of radioactive elements U/Th and stable elements Pb/Os from the $r$-process are found to have a strong correlation.
This correlation is quite robust with respect to astrophysical conditions.
The U/Th-Pb/Os correlation is then applied to provide customized initial abundance ratios U/Th from the observed abundance ratios Pb/Os for six $r$-process enhanced metal-poor stars respectively.
Ages of these six metal-poor stars are predicted by the U/Th chronometer, which are approximately between $11$ and $15$ Gyr.
Their ages are compatible with the cosmic age of 13.8 billion years predicted from the cosmic microwave background radiation.
\end{abstract}



\section{Introduction}\label{Section1}

Age of the universe is one of the most fundamental quantities in cosmology, which is important for understanding the origin and evolution of the Universe~\citep{Gamow1946Phys.Rev.}, the dark matter and dark energy~\citep{Peebles2003Rev.Mod.Phys.}, and the basic theory of gravitation~\citep{Caldwell2009ARNPS}.
Perhaps even more importantly, it is crucial for satisfying human curiosity.
In modern cosmology, the age of the Universe refers to the time elapsed since the Big Bang.
Although humanity's understanding of the cosmic age has gradually deepened over time, debate still exists~\citep{Bennett2013Astrophys.J.Suppl.Ser., PlanckCollaboration2020Astron.Astrophys., Jee2019Science, Chen2019Nat.Astron.}.
Ages of the very metal-poor stars~\citep{Beers2005Nucl.Phys.A} can set lower limits on the cosmic age, as they are believed to be formed at the early epoch of the universe~\citep{Piatti2017Mon.Not.Roy.Astron.Soc.}, and their surface abundances have probably not changed since these stars were formed, except for radioactive decay.
Ages of these stars can thus be determined by the technique of nuclear cosmochronology, which compares the observed present abundances of long-lived radioactive nuclides with the theoretically predicted initial abundances.
This method is independent of cosmological distance and Galactic evolution, which can be used as an independent dating technique for the Universe.

The long half-lives of actinide nuclides $^{232}$Th and $^{238}$U render them ideal candidates as nuclear chronometers for determining the ages of very metal-poor stars.
The present abundances for these nuclides can be obtained from astrophysical observations, while their initial abundances should be given by theoretical simulations of the rapid neutron capture nucleosynthesis process ($r$-process) as actinides can only be produced by this process.
A number of investigations in $r$-process nuclear cosmochronology have been made so far in the literature~\citep{Fowler1960Ann.Phys., Butcher1987thorium, Thielemann1993PR, Cowan1999Astrophys.J., Goriely2001Astron.Astrophys., Schatz2002Astrophys.J., Otsuki2003NewAstron., Niu2009Phys.Rev.C, Kratz2007Astrophys.J., Heng2014JPG, Wu2022Astrophys.J.152, Wu2023Sci.Bull.}.
Most of these works provided zero-decay abundances for the actinide nuclides from $r$-process simulations that are able to reproduce the Solar $r$-process abundances of stable nuclides or inferred for the solar system.
The zero-decay abundances are then utilized for different metal-poor stars.
This is grounded in the assumption that initial $r$-process abundance patterns are the same for different stars, including the sun, namely the universality assumption of $r$-process abundance patterns.
Under this assumption, the uncertainties of nuclear cosmochronology from both nuclear physics inputs and astrophysical conditions have been systematically studied~\citep{Niu2009Phys.Rev.C}, which are quite considerable.

The $r$-process elements of these metal-poor stars came from different $r$-process events, and the astrophysical sites and conditions of these events are probably to be different.
In the 2000s, it had been realized that the universality assumption of $r$-process abundance patterns would not hold for elements $Z\leq 56$ and $Z\geq 75$~\citep{Otsuki2003NewAstron.}.
Nowadays, more and more studies indicate that the $r$-process abundances could be significantly influenced by different astrophysical sites and conditions~\citep{Arnould2007Phys.Rep., Thielemann2011Prog.Part.Nucl.Phys., Kajino2019Prog.Part.Nucl.Phys., Li2019Sci.ChinaPhys.Mech.Astron., Cowan2021Rev.Mod.Phys.}. 
Hence, the universality assumption of $r$-process abundances in the nuclear cosmochronology is probably inappropriate.
Actually, the predicted ages of actinide boost stars under the universality assumption of $r$-process abundances could be unreasonable or even negative~\citep{Schatz2002Astrophys.J., Hill2017Astron.Astrophys.}.
It is thus more reasonable to provide customized initial abundances for each star.
A previous work~\citep{Wu2022Astrophys.J.152} made progress in this direction by constraining customized $r$-process astrophysical condition for each star based on the synchronization requirement of Th/X, U/X, and Th/U chronometers.
It was shown that the idea of synchronizing Th/X, U/X, and Th/U chronometers to provide customized initial abundances for each star works well to reduce the astrophysical uncertainties of the predicted ages, and reasonable ages were obtained for all the considered metal-poor stars, including two actinide boost stars.

In practice, preparing customized initial abundances for each star implies that one needs to determine zero-decay abundances of radioactive elements for each star based on the observed present abundances of both stable and radioactive elements.
In this paper, an $r$-process abundance correlation between the ratio of radioactive elements and the ratio of stable elements is introduced.
It is then applied to provide customized initial abundances for six metal-poor stars respectively, and the ages of these six metal-poor stars are predicted.

\section{$r$-process simulations}\label{Section2}

One should in principle perform $r$-process simulations based on a promising astrophysical scenario.
However, the astrophysical sites responsible for the $r$-process abundances in the very metal-poor stars are still not fully understood, with some proposed scenarios still under debate.
Ejecta from neutron star mergers (NSMs) is supported to be an $r$-process site by the GRB170817 kilonova~\citep{Abbott2017Astrophys.J., Pian2017Nature, Watson2019Nature} associated with gravitational waves from GW170817~\citep{Abbott2017Phys.Rev.Lett.}.
Nevertheless, the binary neutron stars normally take billions of years to collide~\citep{Kobayashi2020Astrophys.J.}, which is too long to match the birth time of the oldest metal-poor stars in the early Universe.
The neutrino driven wind (NDW)~\citep{Woosley1994Astrophys.J.} and the magneto-hydrodynamic jet (MHDJ)~\citep{Nishimura2015Astrophys.J., Nishimura2012Phys.Rev.C} from core-collapse supernova (CCSN), and the outflows from collapsar~\citep{Siegel2019Nature, Nakamura2015Astron.Astrophys., Famiano2020Astrophys.J.} are favored $r$-process site candidates that could have occurred in the early Universe. 
However, there are debates on whether desired high-entropy conditions to produce actinides can occur in the NDW from CCSN~\citep{Fischer2010Astron.Astrophys.}.
The rapid time scale of the MHDJ from CCSN and the outflows from collapsar would lead to the underproduction of isotopic abundances above and below the main $r$-process peaks~\citep{Kajino2019Prog.Part.Nucl.Phys.}, which are different from the observed $r$-process abundances.
Since different scenarios all suffer from some kinds of controversies,  we would perform $r$-process simulations based on proper parametric astrophysical conditions instead of realistic astrophysical scenarios.

In the present study, we perform $r$-process simulations based on a parametric high-entropy wind (HEW) model~\citep{Farouqi2010Astrophys.J.}.
Nucleosynthesis simulations with the HEW model have been extensively studied for long years and are found to reproduce the $r$-process abundances~\citep{Farouqi2010Astrophys.J., Zhao2019Astrophys.J.} well.
This makes the HEW the most popular model for nuclear chronometer studies to generate initial elemental abundances~\citep{Hill2017Astron.Astrophys., Placco2017Astrophys.J., Wu2022Astrophys.J.152, Wu2023Sci.Bull.}.
The $r$-process nucleosynthesis simulations are performed with NucNet tools~\citep{Meyer2013Proc.ofScienceNICXII}.
The nuclear physics inputs are built based on two mass tables respectively, which are WS4~\citep{Wang2014Phys.Lett.B} and DZ28~\citep{Duflo1995Phys.Rev.C}.
Based on the DZ28 and WS4 mass tables, the HEW $r$-process simulations can provide excellent $r$-process abundances for both stable and radioactive nuclides~\citep{Farouqi2010Astrophys.J., Zhao2019Astrophys.J., Wu2022Astrophys.J.152}.
For each set, the nuclear masses are taken from one of these theoretical mass tables when the experimental data~\citep{Wang2012Chin.Phys.C} are unavailable. 
The $(n,\gamma)$ reaction rates are calculated by TALYS-1.9~\citep{Koning2007} with the default settings and parameters except for the nuclear masses.
The $\beta$ decay rates are calculated consistently with the mass model by using the relationship $T_{1/2}\propto1/Q^5$ based on the predictions of the FRDM+QRPA method~\citep{Moeller2003Phys.Rev.C}.
The charged particle reactions and other inputs are sourced from REACLIB2.2 of the JINA Reaclib database~\citep{Cyburt2010Astrophys.J.Suppl.Ser.}.

The evolutions of temperature $T_{9}$~[GK] and density $\rho_{5}$~[$10^{5}$g~cm$^{-3}$] in the HEW model~\citep{Farouqi2010Astrophys.J.} are parameterized as
\begin{align}
  &T_{9}(t) = T_{9}(0)\left( \frac{R_{0}}{R_{0}+V_{\mathrm{exp}}t} \right),\\
  &\rho_{5}(t) = 1.21 \frac{T_{9}^{3}}{S}\left( 1+\frac{7}{4} \frac{T_{9}^2}{T_{9}^{2}+5.3} \right),
\end{align}
where $R_{0}$ is the initial radius of an expanding sphere, $V_{\mathrm{exp}}$ is the expansion velocity of that sphere, and $S$ is entropy in units of $k_{\rm B}$ per baryon.
The total neutrino-wind ejecta consists of various zones with different entropies, which requires a superposition of the contributions from different entropies. 
In the present work, the superposition of entropies from $S=5$ up to $S_{\mathrm{final}}$ (in steps of 5) is made with the weights of inverse entropies~\citep{Farouqi2010Astrophys.J.}.
As a result, the abundance $Y_{\mathrm{sum}}(Z,A)$ of a given nucleus $(Z,A)$ is
\begin{equation}
  Y_{\mathrm{sum}}(Z,A) = \sum_{S=5}^{S_{\mathrm{final}}} \frac{S_{\mathrm{ref}}}{S} Y_{S}(Z,A),
\end{equation}
where $S_{\mathrm{ref}}$ represents an arbitrary reference entropy.
In this work, the electron abundance $Y_e$ and maximum entropy $S_{\mathrm{final}}$ are taken as free parameters, while other parameters are taken as $R_{0} = 130$~km, $V_{\rm exp} = 7500$~km~s$^{-1}$, and $T_{9}(0)=9$.
The dynamical expansion timescale can be defined as the time between initial temperature $T_9=9$ and its decrease to $T_9=3$~\citep{Farouqi2010Astrophys.J.}, which is $\tau_{\rm exp} = \frac{2R_0}{V_{\rm exp}} = 35$~ms in the present calculations.

\section{Results and discussion}\label{Section3}

\subsection{$r$-process abundances}

\begin{figure}[!ht]
	\centering
	\includegraphics[width=0.5\textwidth]{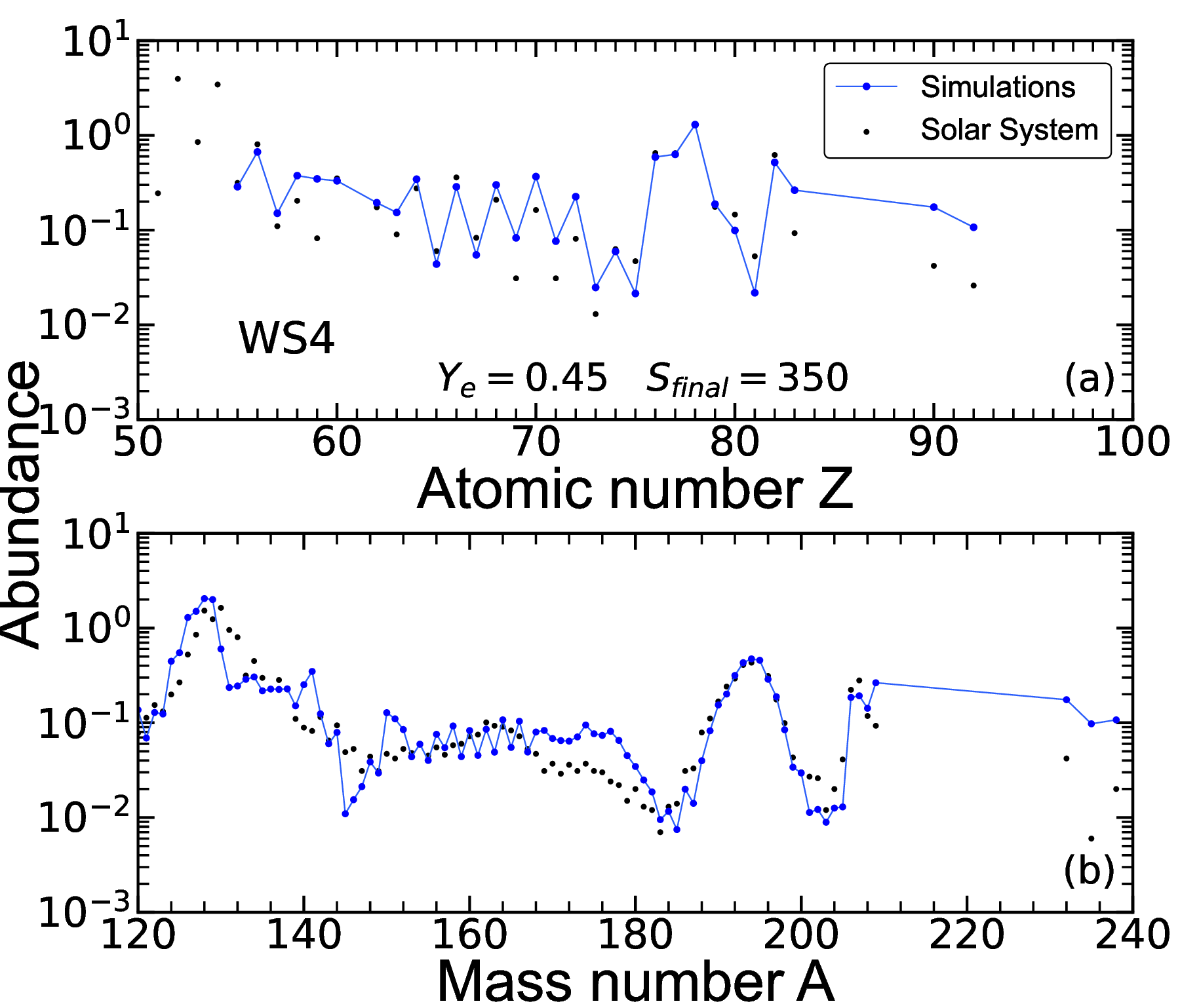}
     \caption{Chemical abundances (a) and nuclide abundances (b) from the $r$-process simulations with WS4 mass model under a specific set of astrophysical condition $(Y_e=0.45, S_{\rm final}=350)$ in comparison with the Solar $r$-process abundances~\citep{Lodders2003Astrophys.J.}.}
     \label{fig:1} 
\end{figure}

A fairly accurate description of $r$-process abundances is a prerequisite for the nuclear cosmochronology study, as the abundances of radioactive and stable nuclides are directly used in the age calculations.
The chemical abundances and nuclide abundances from the $r$-process simulations with WS4 mass model under a specific set of astrophysical condition $(Y_e=0.45, S_{\rm final}=350)$ are depicted in Figure~\ref{fig:1} in comparison with the Solar $r$-process abundances.
An overall agreement between the $r$-process abundances from simulation and the Solar $r$-process abundances can be seen for both Figure~\ref{fig:1} (a) and (b).
In particular, one can observe that the $r$-process simulation based on the WS4 mass model perfectly reproduces the Solar $r$-process abundances around the $A=195$ peak and around $^{208}$Pb.
As will be seen below, this is important for our present studies.

\begin{figure}[!ht]
	\centering
	\includegraphics[width=0.5\textwidth]{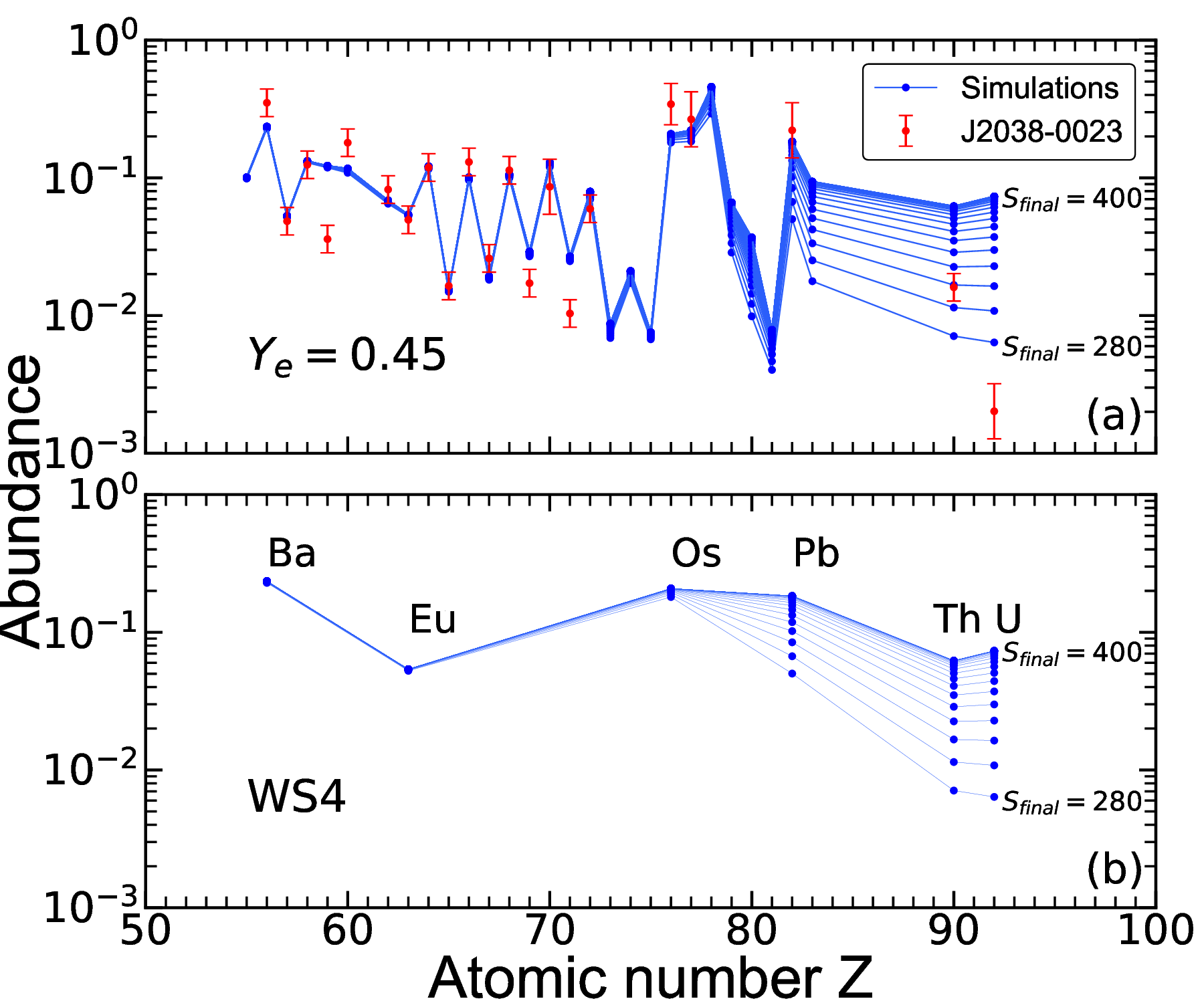}
     \caption{(a) Chemical abundances from the $r$-process simulations with different astrophysical conditions, i.e., fixed electron abundance $Y_e=0.45$ and the entropy $S_{\mathrm{final}}$ varies from $280$ to $400$, in comparison with the observed abundances of the metal-poor star J2038-0023~\citep{Placco2017Astrophys.J.}. The nuclear physics inputs are based on the WS4 mass model~\citep{Wang2014Phys.Lett.B}. (b) Similar with (a), but only the abundances of Ba, Eu, Os, Pb, Th, and U elements are shown.}
     \label{fig:2}
\end{figure}

Figure~\ref{fig:2} (a) shows the chemical abundances from the $r$-process simulations with different astrophysical conditions, i.e., fixed electron abundance $Y_e=0.45$ and the entropy $S_{\mathrm{final}}$ varies from $280$ to $400$, in comparison with the observed abundances of the metal-poor star J2038-0023~\citep{Placco2017Astrophys.J.}.
As can be seen, the abundances of elements with atomic number $Z \le 75$ are barely changed with different astrophysical conditions.
However, for atomic number $Z > 75$, more abundances of elements are produced with larger $S_{\mathrm{final}}$, since they lead to a larger ratio of neutrons to seed nuclei in the environment~\citep{Farouqi2010Astrophys.J.}.
The dependences of the yields for Th and U elements would certainly bring uncertainty to the corresponding nuclear chronometers.

To avoid this uncertainty, one should in principle find a proper way to determine which abundances from different astrophysical conditions to be used in the nuclear cosmochronology.
In a previous work~\citep{Wu2022Astrophys.J.152}, the idea of synchronizing Th/X, U/X, and Th/U chronometers is employed to impose stringent constraints on the astrophysical conditions, which works well to reduce the astrophysical uncertainties of the predicted ages of metal-poor stars.
Here, we notice that the dependence of the $r$-process yields with respect to different astrophysical conditions appear not only for radioactive Th and U elements but also for stable element Pb.
One can see the dependence vary clearly in Figure~\ref{fig:2} (b).
The abundances of the Pb, Th, and U elements increase and decrease simultaneously, which indicates strong correlations among them in the $r$-process.

\subsection{U/Th-Pb/Os correlation}

\begin{figure}[!ht]
	\centering
	\includegraphics[width=0.5\textwidth]{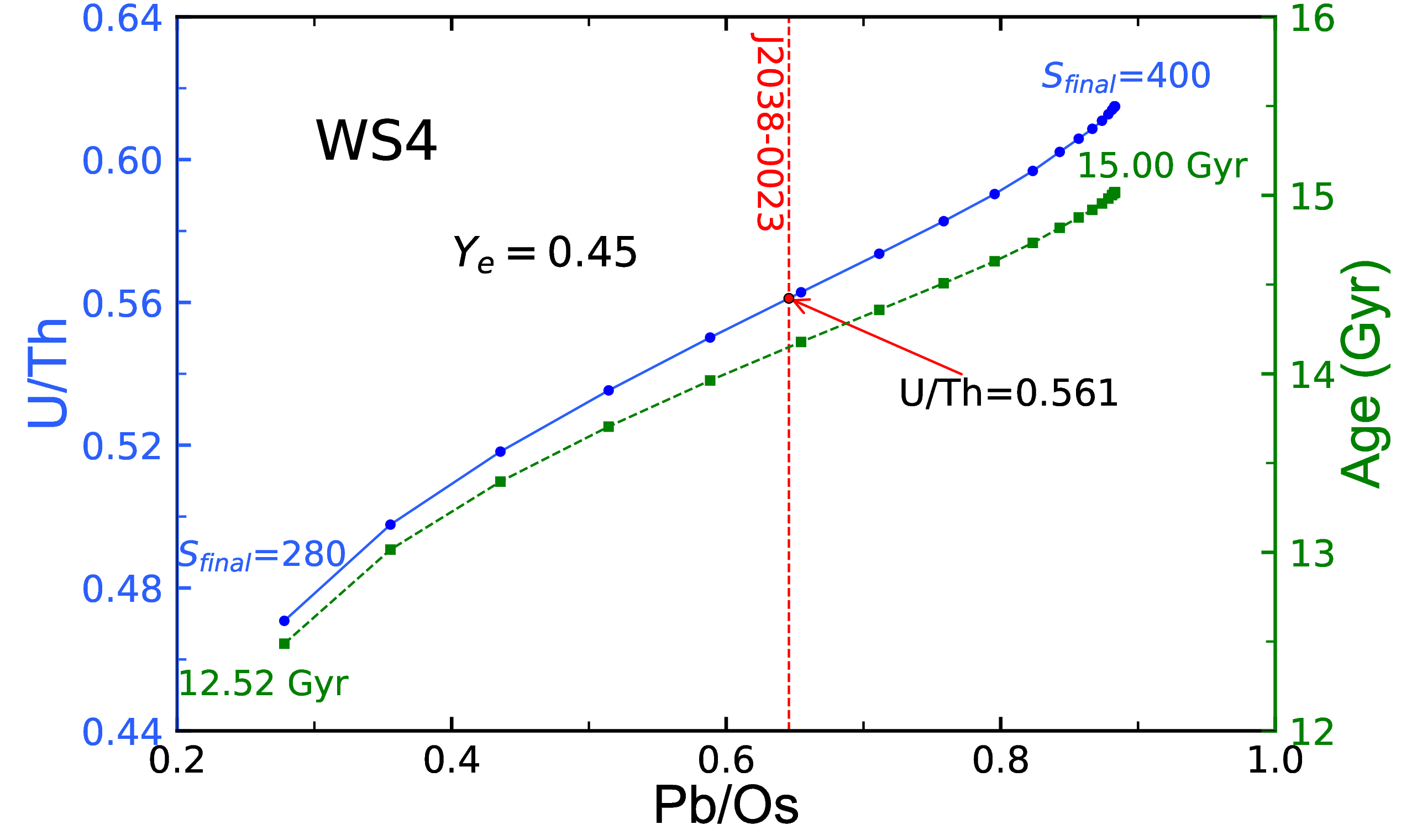}
     \caption{The relation between abundance ratios of U/Th and Pb/Os. 
     The abundances are taken from simulations with electron abundance $Y_e=0.45$ and different entropy $S_{\mathrm{final}}$.
     The ages of the metal-poor star J2038-0023 calculated by adopting the corresponding U/Th ratios are presented by green dash line.
     The vertical line represents the observed Pb/Os ratio of the metal-poor star J2038-0023.}
     \label{fig:3}
\end{figure}

Since no decay occurs, the abundances of stable elements have remained unchanged since they were produced.
As the $r$-process yields of Pb element are sensitive to the astrophysical conditions,
one can in principle use the observed abundances of Pb element to constrain the $r$-process astrophysical conditions, and thus constrain the initial abundances of radioactive Th and U elements.
The relation between abundance ratios of U/Th and Pb/Os is presented in Figure~\ref{fig:3} by taking $Y_e=0.45$ as an example.
The abundance ratio Pb/Os is chosen because Os and Pb are neighboring nuclei in atomic numbers, and the corresponding abundance ratio is less affected by the uncertainties in the overall $r$-process simulation.
As can be seen, there is a strong correlation between U/Th and Pb/Os ratios.
It means that once the ratio of Pb/Os is known, the ratio of U/Th can be predicted correspondingly.
Note that for each specific $S_{\mathrm{final}}$ value, the $r$-process simulation will yield an abundance ratio of U/Th and an abundance ratio of Pb/Os, which contribute to the blue circular points in Figure~\ref{fig:3}. The intermediate cases between these points are obtained by linear interpolation. This gives us a continuous mapping between Pb/Os and U/Th ratios, enabling the determination of a unique U/Th ratio for any observed Pb/Os value.
This would be very useful for the nuclear cosmochronology study, as one can use the observed Pb/Os ratio to constrain the U/Th ratio.
For example, the observed Pb/Os ratio of the metal-poor star J2038-0023 is 0.645, which suggests the initial U/Th ratio to be 0.561 according to the relation shown in Figure~\ref{fig:3}.
Together with the observed U/Th ratio of 0.126, one can then calculate the age of the J2038-0023 star with
\begin{equation}\label{Uthchro}
  \mbox{Age}_{\mbox{Th}/\mbox{U}} = 21.80~{\rm Gyr} [\log(\mbox{U}/\mbox{Th})_{\rm ini.} - \log(\mbox{U}/\mbox{Th})_{\rm obs.}].
\end{equation}
The obtained age of the J2038-0023 star is $14.14$ Gyr.
If the constraint from U/Th-Pb/Os correlation does not apply, all the possible U/Th ratios presented in Figure~\ref{fig:3} could be chosen as the initial abundance ratios, and the estimated ages range from 12.5 to 15.0 Gyr with considerable uncertainty.
Therefore, the application of U/Th-Pb/Os correlation can be in principle valuable for reducing the age-estimation uncertainty in the nuclear cosmochronology study.
It can provide a customized initial abundance ratio of U/Th for each metal-poor star depending on the observed abundance ratio of Pb/Os.

\begin{figure}[!ht]
	\centering
	\includegraphics[width=0.8\textwidth]{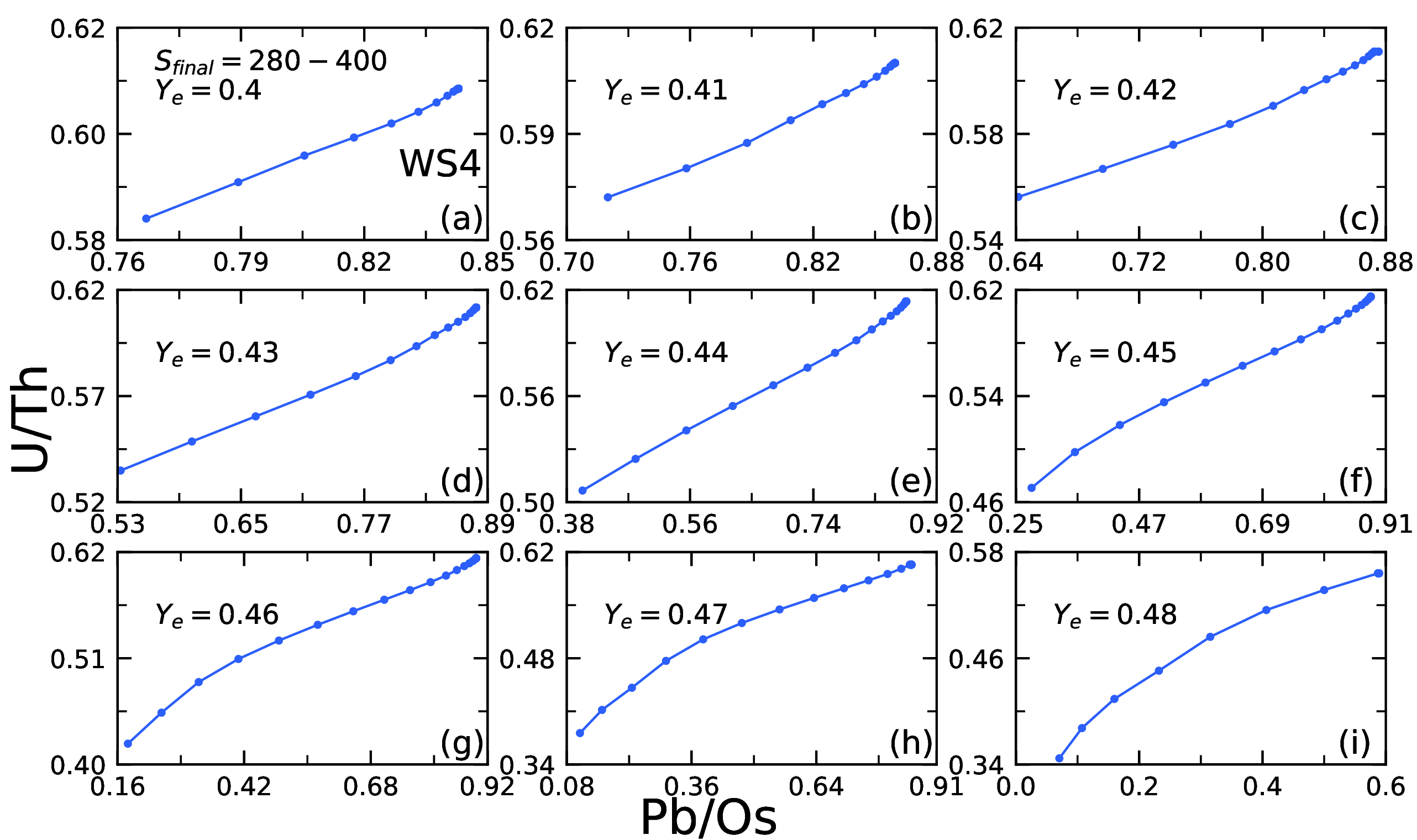}
     \caption{U/Th-Pb/Os correlation under different astrophysical conditions. The electron abundance $Y_e$ varies from 0.40 to 0.48, and the entropy $S_{\mathrm{final}}$ varies from $280$ to $400$. 
     Each subgraph presents the correlation under a specific $Y_e$.}
     \label{fig:4}
\end{figure}

This U/Th-Pb/Os correlation could be affected by the uncertainties arising from astrophysical conditions and nuclear physics inputs in the $r$-process simulations, which need to be addressed.
Figure~\ref{fig:4} (a)-(i) shows the U/Th-Pb/Os correlations under different astrophysical conditions.
One can see from each subgraph of Figure~\ref{fig:4} that the strong U/Th-Pb/Os correlation holds for each $Y_e$, which indicates that this correlation is quite robust with respect to astrophysical conditions.
However, visible differences still exist among the results of different $Y_e$. 
These differences contribute to the uncertainty of this relation, which arises from variations in astrophysical conditions.

\begin{figure}[!ht]
	\centering
	\includegraphics[width=0.5\textwidth]{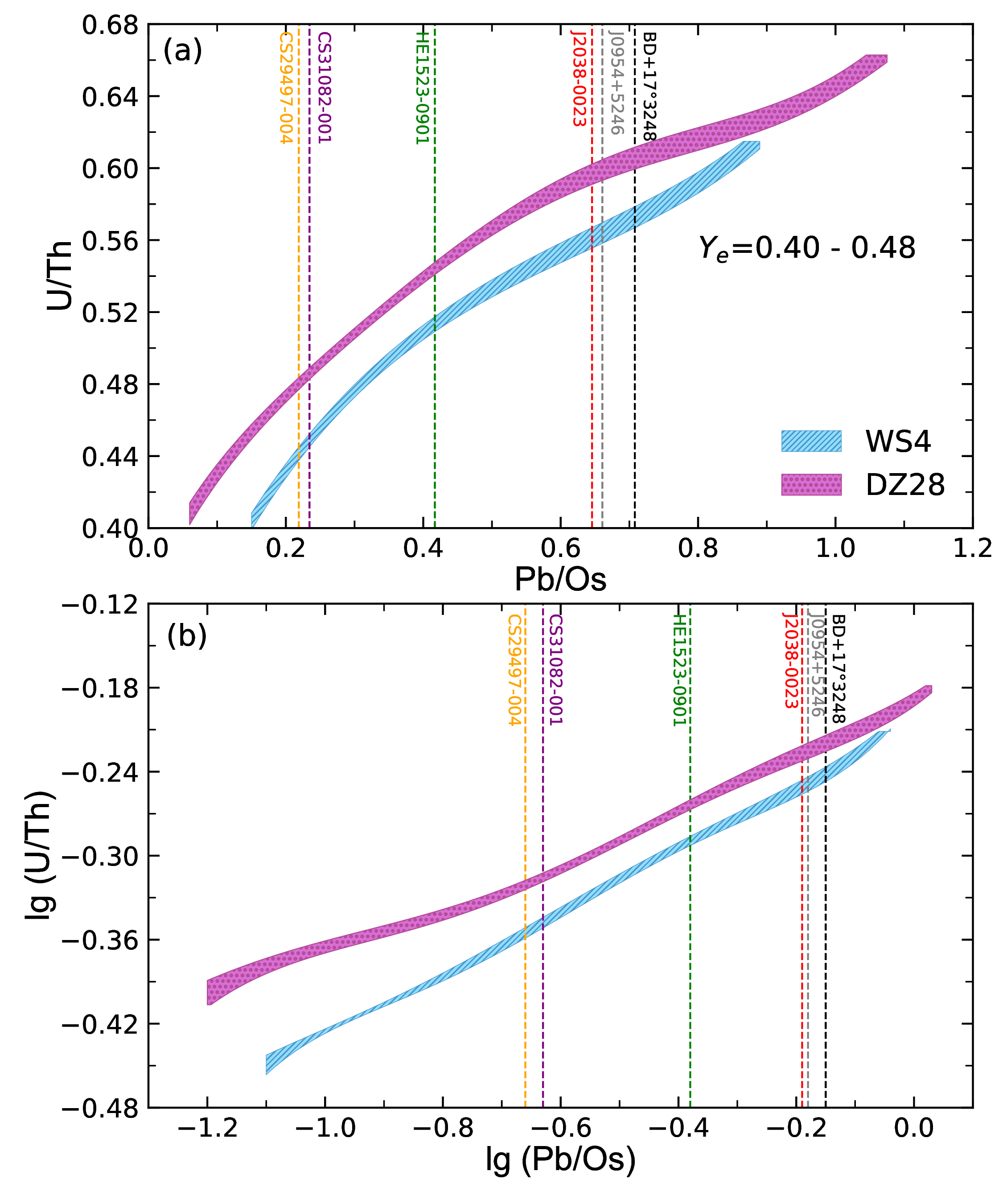}
     \caption{(a) U/Th-Pb/Os correlation under different astrophysical conditions and different sets of nuclear physics inputs. 
     The electron abundance $Y_e$ varies from 0.40 to 0.48, and the entropy $S_{\mathrm{final}}$ varies from $280$ to $400$.
     The two sets of nuclear physics inputs are based on the WS4 and DZ28 mass tables respectively.
     The six vertical lines represent the observed Pb/Os ratio of the metal-poor stars CS31082-001~\citep{Hill2002Astron.Astrophys.}, BD$+$17$^\circ$3248~\citep{Cowan2002Astrophys.J.}, HE1523-0901~\citep{Frebel2007Astrophys.J.}, J2038-0023~\citep{Placco2017Astrophys.J.}, CS29497-004~\citep{Hill2017Astron.Astrophys.}, and J0954+5246~\citep{Holmbeck2018Astrophys.J.}.
     (b) The same as panel (a), but logarithmic coordinates are used.
     }
     \label{fig:5}
\end{figure}

The U/Th-Pb/Os correlations under different astrophysical conditions contribute to a band as shown in Figure~\ref{fig:5}, i.e., light blue band for WS4 and light purple band for DZ28 respectively.
The widths of the bands denote the uncertainties of the U/Th-Pb/Os correlations, which will subsequently contribute to the uncertainties of the ages of metal-poor stars.
One can see that both the bands themselves are quite narrow.
This indicates that the robustness of U/Th-Pb/Os correlation with respect to different astrophysical conditions holds true when different sets of nuclear physics inputs are employed in the $r$-process simulations.

Therefore, it can be reasonably expected that this U/Th-Pb/Os correlation will also hold true with more accurate nuclear physics inputs from both experiments and theories in the future. 
Efforts in two directions would be important to provide accurate $r$-process nuclear physics inputs. 
One direction is to build $r$-process inputs based on reliable latest microscopic nuclear models, e.g., building $r$-process inputs based on the on-going DRHBc mass table project~\citep{Zhang2022Atom.DataNucl.DataTables, Pan2022Phys.Rev.C, Guo2024ADNDT, Jiang2024PLB, Wu2024Phys.Rev.C, Guo2024PRC, Zhang2025AAPPSBulletin}. 
Meanwhile, the other direction involves enhancing these inputs through novel machine learning techniques or empirical formulae, e.g., building machine-learning nuclear mass models~\citep{Niu2018Phys.Lett.B, Wu2020Phys.Rev.C051301, Wu2021Phys.Lett.B, Niu2022Phys.Rev.C, Wu2022Phys.Lett.B137394, Wu2024PRC_AKRR} and empirical $\beta$-decay formulae~\citep{Zhou2017Sci.ChinaPhys.Mech.Astron., Tian2025CPC}.
It is also important to note that these two uncertainty bands are at variance with each other, corresponding to the well-known $r$-process uncertainties from nuclear physics inputs~\citep{Mumpower2016Prog.Part.Nucl.Phys., Jiang2021Astrophys.J.}.
This means that the detailed U/Th-Pb/Os correlation is depended on nuclear physics inputs.
The difference as shown in Figure~\ref{fig:5} stands for the nuclear physics uncertainty of the U/Th-Pb/Os correlation.
As can be seen from Figure~\ref{fig:5} (b), for every increase of 1.0 in lg(Pb/Os), lg(U/Th) increases by 0.2. 
This also implies that an observation error of 1.0 in lg(Pb/Os) would lead to an uncertainty of 0.2 in initial lg(U/Th), which would propagate to the uncertainty of age determinations.

\subsection{Ages of the metal-poor stars}

The uranium abundance has been observed in six metal-poor stars, i.e., CS31082-001~\citep{Hill2002Astron.Astrophys.}, BD$+$17$^\circ$3248~\citep{Cowan2002Astrophys.J.}, HE1523-0901~\citep{Frebel2007Astrophys.J.}, J2038-0023~\citep{Placco2017Astrophys.J.}, CS29497-004~\citep{Hill2017Astron.Astrophys.}, and J0954+5246~\citep{Holmbeck2018Astrophys.J.}. 
Therefore, the U/Th chronometer can be applied to date the ages of these six metal-poor stars.
The observed Pb/Os ratios are different among these six stars, and, thus, the corresponding U/Th ratios determining by the U/Th-Pb/Os correlation would be different.
As shown in Figure~\ref{fig:5}, different metal-poor stars with Pb/Os ratios have different intersections with the U/Th-Pb/Os correlation bands as shown in Figure~\ref{fig:5}.
One can thus provide a customized initial abundance ratio range of U/Th for each metal-poor star depending on the observed abundance ratio of Pb/Os of the corresponding stars.

\begin{table*}[h]
\centering
\caption{The ages and associated errors of six metal-poor stars calculated by the U/Th chronometer with initial abundances obtained from the U/Th-Pb/Os correlation.
The observed abundance ratios of Pb/Os and U/Th are also presented.
The units of ages and associated Uncertainties are Gyr.
The column ``Obs. Uncertainty" represents the age uncertainties caused by the observed uncertainties of both U/Th ratios and Pb/Os ratios. The uncertainty in Pb/Os ratios propagates to the age determinations by affecting the initial U/Th ratios via the U/Th-Pb/Os correlation relationship.
}
\label{tab:1}
{\small
\setlength{\tabcolsep}{4pt} 
\begin{tabular}{lccccccccc}
\hline
\multicolumn{4}{c}{Stars} & \multicolumn{3}{c}{WS4} & \multicolumn{3}{c}{DZ28}\\  
\cmidrule(lr){1-4} \cmidrule(lr){5-7} \cmidrule(lr){8-10} 
Name & {$\left(\frac{\rm Pb}{\rm Os}\right)_{\rm obs.}$} & {$\left(\frac{\rm U}{\rm Th}\right)_{\rm obs.}$} & {Obs. Uncertainty} & Rough Age & Revised Age & Uncertainty & Rough Age & Revised Age & Uncertainty  \\
\midrule
\midrule  
J2038-0023    & 0.65 & 0.13 & 1.88 & 14.14 & 13.85 & 0.09 & 14.71 & 14.44 & 0.08 \\
CS31082-001   & 0.23 & 0.11 & 1.44 & 12.86 & 12.19 & 0.10 & 13.65 & 13.19 & 0.08 \\
BD+17°3248    & 0.71 & 0.15 & 2.69 & 12.59 & 12.47 & 0.10 & 13.12 & 13.03 & 0.09 \\
HE1523-0901   & 0.42 & 0.14 & 1.50 & 12.43 & 12.05 & 0.07 & 12.99 & 12.59 & 0.05 \\
CS29497-004   & 0.22 & 0.09 & 2.90 & 14.87 & 14.33 & 0.10 & 15.70 & 15.34 & 0.08 \\
J0954+5246    & 0.66 & 0.15 & 1.99 & 12.44 & 12.03 & 0.09 & 13.00 & 12.62 & 0.07 \\
\bottomrule  
\end{tabular}
}
\end{table*}

The ages of these metal-poor stars can thus be calculated with the U/Th chronometer with initial abundances obtained from the U/Th-Pb/Os correlation using Equation~\eqref{Uthchro}. 
Ages of these six stars are listed in Table~\ref{tab:1}, i.e., columns of rough age.
However, it should be noted that the radioactive nuclides $^{238}$U and $^{232}$Th will eventually decay to the stable nuclides $^{206}$Pb and $^{208}$Pb, respectively.
This means that the observed abundance of the Pb element includes some parts that have decayed from radioactive elements Th and U.
One should therefore make corrections to the Pb abundances, which will affect the Pb/Os ratios and then the related U/Th ratios for the six stars.
Note that other radioactive isotopes heavier than Pb have relatively short half-lives, which had already decayed away in the late stage of the $r$-process.
The rough ages in the first round of estimations together with the present observed $^{238}$U and $^{232}$Th abundances are used to calculate the amount of $^{238}$U and $^{232}$Th isotopes that decay to Pb element for each star.
This part of abundance is then subtracted from the observed Pb abundance.
With the corrected Pb/Os ratios, the initial U/Th ratios are then updated from the U/Th-Pb/Os correction.
The revised ages are also presented in Table~\ref{tab:1}.

Ages of these six stars are predicted to be around $11$ to $15$ Gyr, which are consistent with the cosmic age $13.8$ Gyr predicted from the microwave background radiation~\citep{PlanckCollaboration2020Astron.Astrophys.}, if the uncertainties from nuclear physics inputs and abundance observations are considered.
However, some of these metal-poor stars are older than the cosmic age $11.4$ Gyr predicted by the recent measurement of Hubble constant from angular diameter distances to two gravitational lenses~\citep{Jee2019Science}.
It should be noted that the ages of two actinide boost stars CS31082-001 and J0954+5246 are also reasonablely predicted by the new recipe introduced in the present work.
This indicates that the U/Th-Pb/Os correlation can cover metal-poor stars whose surface abundances originate from extreme $r$-process conditions.
These results demonstrate the effectiveness of using the U/Th-Pb/Os correlation for nuclear cosmochronology studies.
Furthermore, the ages of these six metal-poor stars predicted in the present work are all consistent with the ages estimated by the Th-U-X chronometer~\citep{Wu2022Astrophys.J.152} within uncertainties.
As can be seen in Table~\ref{tab:1}, the uncertainties associated with observed elemental abundances in metal-poor stars are rather large. They are expected to be reduced by the efforts including the acquisition of high resolution and low-noise spectra of the stars, the construction of realistic model stellar atmospheres, the analysis of the spectra with few limiting simplifications, and the improvement in basic atomic and molecular data~\citep{Sneden2009Astrophys.J.Suppl.Ser.}.

The present recipe of nuclear cosmochronology that utilizes the relation of U/Th-Pb/Os abundance ratios has the following advantages:
\begin{enumerate}
  \item[(i)] Goes beyond the universality assumption of $r$-process abundance patterns.
  The $r$-process elements of different metal-poor stars came from different $r$-process events, and the corresponding astrophysical sites and conditions of these events could be very different.
  According to present-day observations and simulations, the $r$-process abundances could be significantly influenced by different astrophysical sites and conditions.
  Therefore, the universality assumption of $r$-process abundances in the nuclear cosmochronology is probably inappropriate, star-specific initial abundances should be derived for each star instead.
  Based on the U/Th-Pb/Os correlation, the present work provides customized initial abundance ratios U/Th for six metal-poor stars respectively, according to the observed abundance ratio of Pb/Os of the corresponding stars.
  \item[(ii)] Reduce age-estimation uncertainties via stable-element constraints.
  Since no decay occurs, the abundances of stable elements of metal-poor stars remained mostly unchanged after their productions.
  Therefore, one can use the observed abundances of stable elements as the initial ones to constrain the $r$-process.
  Based on the U/Th-Pb/Os correlation, the present work constrains the initial U/Th abundance ratio based on the observed Pb/Os ratio for each star.
  This reduces the uncertainties of the initial U/Th abundance ratio and thus the corresponding age-estimation uncertainties by nuclear cosmochronology.
\end{enumerate}

\section{Summary}

The abundance ratios of radioactive elements U/Th and stable elements Pb/Os from the $r$-process are found to have a strong correlation.
Leveraging this correlation, one can then constrain the initial U/Th abundance ratio based on the observed Pb/Os ratio, which offers a novel recipe of nuclear cosmochronology.
The dependence of this U/Th-Pb/Os correlation on $r$-process uncertainties from astrophysics and nuclear physics is studied.
It is found that the U/Th-Pb/Os correlation is quite robust with respect to astrophysical conditions.
However, its dependence on nuclear physics inputs is still significant.
This U/Th-Pb/Os correlation is then applied to provide customized initial abundances of radioactive elements for six metal-poor stars respectively, which include two actinide boost $r$-process enhanced metal-poor stars CS31082-001 and J0954+5246. 
Ages of these six metal-poor stars are predicted to be approximately between $11$ and $15$ Gyr, which are consistent with the cosmic age $13.8$ Gyr predicted from the microwave background radiation~\citep{PlanckCollaboration2020Astron.Astrophys.} if the uncertainties are considered.
However, some of these metal-poor stars are older than the cosmic age $11.4$ Gyr predicted by the recent measurement of Hubble constant from angular diameter distances to two gravitational lenses~\citep{Jee2019Science}, which means that this younger cosmic age is not supported by our studies.
These results demonstrate the effectiveness of using the U/Th-Pb/Os correlation for nuclear cosmochronology studies.
In the future, with the implementation of $r$-process-enhanced star survey projects, such as the Hamburg/ESO $r$-process-enriched star search project in Germany~\citep{Christlieb2004Astron.Astrophys.}, the observational project of the $r$-process alliance led by the United States~\citep{Hansen2018Astrophys.J.}, and the LAMOST $r$-process-enriched star search project in China~\citep{Chen2021Res.Astron.Astrophys.}, an increasing number of $r$-process-enriched extremely metal-poor stars with observed abundances of the U element will be discovered.
The U/Th-Pb/Os correlation can be applied to determine the ages of these stars, thus providing more accurate and stringent constraints on the age of the universe.

\section*{acknowledgments}
X.H.W. thanks Zhongming Niu for helpful discussions.
This work was supported by the National Natural Science Foundation of China (Grants No. 12405134, No. 12435006, No. 12141501, No. 12475117), the National Key R\&D Program of China (Contract No. 2024YFE0109803), the State Key Laboratory of Nuclear Physics and Technology, Peking University (Grant No. NPT2023ZX03), and the National Key Laboratory of Neutron Science and Technology (Grant No. NST202401016).


\bibliography{paper}{}

\begin{thebibliography}{}
\expandafter\ifx\csname natexlab\endcsname\relax\def\natexlab#1{#1}\fi
\providecommand{\url}[1]{\href{#1}{#1}}
\providecommand{\dodoi}[1]{doi:~\href{http://doi.org/#1}{\nolinkurl{#1}}}
\providecommand{\doeprint}[1]{\href{http://ascl.net/#1}{\nolinkurl{http://ascl.net/#1}}}
\providecommand{\doarXiv}[1]{\href{https://arxiv.org/abs/#1}{\nolinkurl{https://arxiv.org/abs/#1}}}

\bibitem[{Abbott {et~al.}(2017{\natexlab{a}})Abbott, Abbott, Abbott, Acernese, Ackley, Adams, Adams, Addesso, Adhikari, Adya, Affeldt, Afrough, Agarwal, Agathos, Agatsuma, Aggarwal, Aguiar, Aiello, Ain, Ajith, Allen, Allen, Allocca, Altin, {et~al.}}]{Abbott2017Astrophys.J.}
Abbott, B.~P., Abbott, R., Abbott, T.~D., {et~al.} 2017{\natexlab{a}}, Astrophys. J., 848, L12, \dodoi{10.3847/2041-8213/aa91c9}

\bibitem[{Abbott {et~al.}(2017{\natexlab{b}})Abbott, Abbott, Abbott, Acernese, Ackley, Adams, Adams, Addesso, Adhikari, Adya, Affeldt, Afrough, Agarwal, Agathos, Agatsuma, Aggarwal, Aguiar, Aiello, Ain, Ajith, Allen, Allen, Allocca, Altin, Amato, Ananyeva, Anderson, Anderson, Angelova, Antier, Appert, Arai, Araya, Areeda, Arnaud, Arun, Ascenzi, Ashton, Ast, Aston, Astone, Atallah, Aufmuth, Aulbert, AultONeal, Austin, Avila-Alvarez, Babak, Bacon, Bader, Bae, Bailes, Baker, Baldaccini, Ballardin, Ballmer, Banagiri, Barayoga, Barclay, Barish, Barker, Barkett, Barone, Barr, Barsotti, Barsuglia, Barta, Barthelmy, Bartlett, Bartos, Bassiri, Basti, Batch, Bawaj, Bayley, Bazzan, B\'ecsy, Beer, Bejger, Belahcene, Bell, Berger, Bergmann, Bernuzzi, Bero, Berry, Bersanetti, Bertolini, Betzwieser, Bhagwat, Bhandare, Bilenko, Billingsley, Billman, Birch, Birney, Birnholtz, Biscans, Biscoveanu, Bisht, Bitossi, Biwer, Bizouard, Blackburn, Blackman, Blair, Blair, Blair, Bloemen, Bock, Bode, Boer, Bogaert, Bohe, Bondu,
  Bonilla, Bonnand, Boom, Bork, Boschi, Bose, Bossie, Bouffanais, Bozzi, Bradaschia, Brady, Branchesi, Brau, Briant, Brillet, Brinkmann, Brisson, Brockill, Broida, Brooks, Brown, Brown, Brunett, Buchanan, Buikema, Bulik, Bulten, Buonanno, Buskulic, Buy, Byer, Cabero, Cadonati, Cagnoli, Cahillane, Calder\'on~Bustillo, Callister, Calloni, Camp, Canepa, Canizares, Cannon, Cao, Cao, Capano, Capocasa, Carbognani, Caride, Carney, Carullo, Casanueva~Diaz, Casentini, Caudill, Cavagli\`a, Cavalier, Cavalieri, Cella, Cepeda, Cerd\'a-Dur\'an, Cerretani, Cesarini, Chamberlin, Chan, Chao, Charlton, Chase, Chassande-Mottin, Chatterjee, Chatziioannou, Cheeseboro, Chen, Chen, Chen, Cheng, Chia, Chincarini, Chiummo, Chmiel, Cho, Cho, Chow, Christensen, Chu, Chua, Chua, Chung, Chung, Ciani, Ciolfi, Cirelli, Cirone, Clara, Clark, Clearwater, Cleva, Cocchieri, Coccia, Cohadon, Cohen, Colla, Collette, Cominsky, Constancio, Conti, Cooper, Corban, Corbitt, Cordero-Carri\'on, Corley, Cornish, Corsi, Cortese, Costa, Coughlin,
  Coughlin, Coulon, Countryman, Couvares, Covas, Cowan, Coward, Cowart, Coyne, Coyne, Creighton, Creighton, Cripe, Crowder, Cullen, Cumming, Cunningham, Cuoco, Dal~Canton, D\'alya, Danilishin, D'Antonio, Danzmann, Dasgupta, Da~Silva~Costa, Dattilo, Dave, Davier, Davis, Daw, Day, De, DeBra, Degallaix, De~Laurentis, Del\'eglise, Del~Pozzo, Demos, Denker, Dent, De~Pietri, Dergachev, De~Rosa, DeRosa, De~Rossi, DeSalvo, de~Varona, Devenson, Dhurandhar, D\'{\i}az, Dietrich, Di~Fiore, Di~Giovanni, Di~Girolamo, Di~Lieto, Di~Pace, Di~Palma, Di~Renzo, Doctor, Dolique, Donovan, Dooley, Doravari, Dorrington, Douglas, Dovale~\'Alvarez, Downes, Drago, Dreissigacker, Driggers, Du, Ducrot, Dudi, Dupej, Dwyer, Edo, Edwards, Effler, Eggenstein, Ehrens, Eichholz, Eikenberry, Eisenstein, Essick, Estevez, Etienne, Etzel, Evans, Evans, Factourovich, Fafone, Fair, Fairhurst, Fan, Farinon, Farr, Farr, Fauchon-Jones, Favata, Fays, Fee, Fehrmann, Feicht, Fejer, Fernandez-Galiana, Ferrante, Ferreira, Ferrini, Fidecaro, Finstad, Fiori,
  Fiorucci, Fishbach, Fisher, Fitz-Axen, Flaminio, Fletcher, Fong, Font, Forsyth, Forsyth, Fournier, Frasca, Frasconi, Frei, Freise, Frey, Frey, Fries, Fritschel, Frolov, Fulda, Fyffe, Gabbard, Gadre, Gaebel, Gair, Gammaitoni, Ganija, Gaonkar, Garcia-Quiros, Garufi, Gateley, Gaudio, Gaur, Gayathri, Gehrels, Gemme, Genin, Gennai, George, George, Gergely, Germain, Ghonge, Ghosh, Ghosh, Ghosh, Giaime, Giardina, Giazotto, Gill, Glover, Goetz, Goetz, Gomes, Goncharov, Gonz\'alez, Gonzalez~Castro, Gopakumar, Gorodetsky, Gossan, Gosselin, Gouaty, Grado, Graef, Granata, Grant, Gras, Gray, Greco, Green, Gretarsson, Groot, Grote, Grunewald, Gruning, Guidi, Guo, Gupta, Gupta, Gushwa, Gustafson, Gustafson, Halim, Hall, Hall, Hamilton, Hammond, Haney, Hanke, Hanks, Hanna, Hannam, Hannuksela, Hanson, Hardwick, Harms, Harry, Harry, Hart, Haster, Haughian, Healy, Heidmann, Heintze, Heitmann, Hello, Hemming, Hendry, Heng, Hennig, Heptonstall, Heurs, Hild, Hinderer, Ho, Hoak, Hofman, Holt, Holz, Hopkins, Horst, Hough, Houston,
  Howell, Hreibi, Hu, Huerta, Huet, Hughey, Husa, Huttner, Huynh-Dinh, Indik, Inta, Intini, Isa, Isac, Isi, Iyer, Izumi, Jacqmin, Jani, Jaranowski, Jawahar, Jim\'enez-Forteza, Johnson, Johnson-McDaniel, Jones, Jones, Jonker, Ju, Junker, Kalaghatgi, Kalogera, Kamai, Kandhasamy, Kang, Kanner, Kapadia, Karki, Karvinen, Kasprzack, Kastaun, Katolik, Katsavounidis, Katzman, Kaufer, Kawabe, K\'ef\'elian, Keitel, Kemball, Kennedy, Kent, Key, Khalili, Khan, Khan, Khan, Khazanov, Kijbunchoo, Kim, Kim, Kim, Kim, Kim, Kim, Kimbrell, King, King, Kinley-Hanlon, Kirchhoff, Kissel, Kleybolte, Klimenko, Knowles, Koch, Koehlenbeck, Koley, Kondrashov, Kontos, Korobko, Korth, Kowalska, Kozak, Kr\"amer, Kringel, Krishnan, Kr\'olak, Kuehn, Kumar, Kumar, Kumar, Kuo, Kutynia, Kwang, Lackey, Lai, Landry, Lang, Lange, Lantz, Lanza, Larson, Lartaux-Vollard, Lasky, Laxen, Lazzarini, Lazzaro, Leaci, Leavey, Lee, Lee, Lee, Lee, Lee, Lehmann, Lenon, Leon, Leonardi, Leroy, Letendre, Levin, Li, Linker, Littenberg, Liu, Liu, Lo, Lockerbie,
  London, Lord, Lorenzini, Loriette, Lormand, Losurdo, Lough, Lousto, Lovelace, L\"uck, Lumaca, Lundgren, Lynch, Ma, Macas, Macfoy, Machenschalk, MacInnis, Macleod, Maga\~na Hernandez, Maga\~na Sandoval, Maga\~na Zertuche, Magee, Majorana, Maksimovic, Man, Mandic, Mangano, Mansell, Manske, Mantovani, Marchesoni, Marion, M\'arka, M\'arka, Markakis, Markosyan, Markowitz, Maros, Marquina, Marsh, Martelli, Martellini, Martin, Martin, Martynov, Marx, Mason, Massera, Masserot, Massinger, Masso-Reid, Mastrogiovanni, Matas, Matichard, Matone, Mavalvala, Mazumder, McCarthy, McClelland, McCormick, McCuller, McGuire, McIntyre, McIver, McManus, McNeill, McRae, McWilliams, Meacher, Meadors, Mehmet, Meidam, Mejuto-Villa, Melatos, Mendell, Mercer, Merilh, Merzougui, Meshkov, Messenger, Messick, Metzdorff, Meyers, Miao, Michel, Middleton, Mikhailov, Milano, Miller, Miller, Miller, Millhouse, Milovich-Goff, Minazzoli, Minenkov, Ming, Mishra, Mitra, Mitrofanov, Mitselmakher, Mittleman, Moffa, Moggi, Mogushi, Mohan, Mohapatra,
  Molina, Montani, Moore, Moraru, Moreno, Morisaki, Morriss, Mours, Mow-Lowry, Mueller, Muir, Mukherjee, Mukherjee, Mukherjee, Mukund, Mullavey, Munch, Mu\~niz, Muratore, Murray, Nagar, Napier, Nardecchia, Naticchioni, Nayak, Neilson, Nelemans, Nelson, Nery, Neunzert, Nevin, Newport, Newton, Ng, Nguyen, Nguyen, Nichols, Nielsen, Nissanke, Nitz, Noack, Nocera, Nolting, North, Nuttall, Oberling, O'Dea, Ogin, Oh, Oh, Ohme, Okada, Oliver, Oppermann, Oram, O'Reilly, Ormiston, Ortega, O'Shaughnessy, Ossokine, Ottaway, Overmier, Owen, Pace, Page, Page, Pai, Pai, Palamos, Palashov, Palomba, Pal-Singh, Pan, Pan, Pang, Pang, Pankow, Pannarale, Pant, Paoletti, Paoli, Papa, Parida, Parker, Pascucci, Pasqualetti, Passaquieti, Passuello, Patil, Patricelli, Pearlstone, Pedraza, Pedurand, Pekowsky, Pele, Penn, Perez, Perreca, Perri, Pfeiffer, Phelps, Piccinni, Pichot, Piergiovanni, Pierro, Pillant, Pinard, Pinto, Pirello, Pitkin, Poe, Poggiani, Popolizio, Porter, Post, Powell, Prasad, Pratt, Pratten, Predoi, Prestegard,
  Prijatelj, Principe, Privitera, Prix, Prodi, Prokhorov, Puncken, Punturo, Puppo, P\"urrer, Qi, Quetschke, Quintero, Quitzow-James, Raab, Rabeling, Radkins, Raffai, Raja, Rajan, Rajbhandari, Rakhmanov, Ramirez, Ramos-Buades, Rapagnani, Raymond, Razzano, Read, Regimbau, Rei, Reid, Reitze, Ren, Reyes, Ricci, Ricker, Rieger, Riles, Rizzo, Robertson, Robie, Robinet, Rocchi, Rolland, Rollins, Roma, Romano, Romano, Romel, Romie, Rosi\ifmmode~\acute{n}\else \'{n}\fi{}ska, Ross, Rowan, R\"udiger, Ruggi, Rutins, Ryan, Sachdev, Sadecki, Sadeghian, Sakellariadou, Salconi, Saleem, Salemi, Samajdar, Sammut, Sampson, Sanchez, Sanchez, Sanchis-Gual, Sandberg, Sanders, Sassolas, Sathyaprakash, Saulson, Sauter, Savage, Sawadsky, Schale, Scheel, Scheuer, Schmidt, Schmidt, Schnabel, Schofield, Sch\"onbeck, Schreiber, Schuette, Schulte, Schutz, Schwalbe, Scott, Scott, Seidel, Sellers, Sengupta, Sentenac, Sequino, Sergeev, Shaddock, Shaffer, Shah, Shahriar, Shaner, Shao, Shapiro, Shawhan, Sheperd, Shoemaker, Shoemaker, Siellez,
  Siemens, Sieniawska, Sigg, Silva, Singer, Singh, Singhal, Sintes, Slagmolen, Smith, Smith, Smith, Somala, Son, Sonnenberg, Sorazu, Sorrentino, Souradeep, Spencer, Srivastava, Staats, Staley, Steinke, Steinlechner, Steinlechner, Steinmeyer, Stevenson, Stone, Stops, Strain, Stratta, Strigin, Strunk, Sturani, Stuver, Summerscales, Sun, Sunil, Suresh, Sutton, Swinkels, Szczepa\ifmmode~\acute{n}\else \'{n}\fi{}czyk, Tacca, Tait, Talbot, Talukder, Tanner, T\'apai, Taracchini, Tasson, Taylor, Taylor, Tewari, Theeg, Thies, Thomas, Thomas, Thomas, Thorne, Thorne, Thrane, Tiwari, Tiwari, Tokmakov, Toland, Tonelli, Tornasi, Torres-Forn\'e, Torrie, T\"oyr\"a, Travasso, Traylor, Trinastic, Tringali, Trozzo, Tsang, Tse, Tso, Tsukada, Tsuna, Tuyenbayev, Ueno, Ugolini, Unnikrishnan, Urban, Usman, Vahlbruch, Vajente, Valdes, Vallisneri, van Bakel, van Beuzekom, van~den Brand, Van Den~Broeck, Vander-Hyde, van~der Schaaf, van Heijningen, van Veggel, Vardaro, Varma, Vass, Vas\'uth, Vecchio, Vedovato, Veitch, Veitch,
  Venkateswara, Venugopalan, Verkindt, Vetrano, Vicer\'e, Viets, Vinciguerra, Vine, Vinet, Vitale, Vo, Vocca, Vorvick, Vyatchanin, Wade, Wade, Wade, Walet, Walker, Wallace, Walsh, Wang, Wang, Wang, Wang, Wang, Ward, Warner, Was, Watchi, Weaver, Wei, Weinert, Weinstein, Weiss, Wen, Wessel, We\ss{}els, Westerweck, Westphal, Wette, Whelan, Whitcomb, Whiting, Whittle, Wilken, Williams, Williams, Williamson, Willis, Willke, Wimmer, Winkler, Wipf, Wittel, Woan, Woehler, Wofford, Wong, Worden, Wright, Wu, Wysocki, Xiao, Yamamoto, Yancey, Yang, Yap, Yazback, Yu, Yu, Yvert, Zadro\ifmmode~\dot{z}\else \.{z}\fi{}ny, Zanolin, Zelenova, Zendri, Zevin, Zhang, Zhang, Zhang, Zhang, Zhao, Zhou, Zhou, Zhu, Zhu, Zimmerman, Zucker, \& Zweizig}]{Abbott2017Phys.Rev.Lett.}
---. 2017{\natexlab{b}}, Phys. Rev. Lett., 119, 161101, \dodoi{10.1103/PhysRevLett.119.161101}

\bibitem[{Aghanim {et~al.}(2020)Aghanim, {Akrami, Y.}, {Ashdown, M.}, {Aumont, J.}, {Baccigalupi, C.}, {Ballardini, M.}, {Banday, A. J.}, {Barreiro, R. B.}, {Bartolo, N.}, {Basak, S.}, {Battye, R.}, {Benabed, K.}, {Bernard, J.-P.}, {Bersanelli, M.}, {Bielewicz, P.}, {Bock, J. J.}, {Bond, J. R.}, {Borrill, J.}, {Bouchet, F. R.}, {Boulanger, F.}, {Bucher, M.}, {Burigana, C.}, {Butler, R. C.}, {Calabrese, E.}, {Cardoso, J.-F.}, {Carron, J.}, {Challinor, A.}, {Chiang, H. C.}, {Chluba, J.}, {Colombo, L. P. L.}, {Combet, C.}, {Contreras, D.}, {Crill, B. P.}, {Cuttaia, F.}, {de Bernardis, P.}, {de Zotti, G.}, {Delabrouille, J.}, {Delouis, J.-M.}, {Di Valentino, E.}, {Diego, J. M.}, {Doré, O.}, {Douspis, M.}, {Ducout, A.}, {Dupac, X.}, {Dusini, S.}, {Efstathiou, G.}, {Elsner, F.}, {Enßlin, T. A.}, {Eriksen, H. K.}, {Fantaye, Y.}, {Farhang, M.}, {Fergusson, J.}, {Fernandez-Cobos, R.}, {Finelli, F.}, {Forastieri, F.}, {Frailis, M.}, {Fraisse, A. A.}, {Franceschi, E.}, {Frolov, A.}, {Galeotta, S.}, {Galli, S.},
  {Ganga, K.}, {Génova-Santos, R. T.}, {Gerbino, M.}, {Ghosh, T.}, {González-Nuevo, J.}, {Górski, K. M.}, {Gratton, S.}, {Gruppuso, A.}, {Gudmundsson, J. E.}, {Hamann, J.}, {Handley, W.}, {Hansen, F. K.}, {Herranz, D.}, {Hildebrandt, S. R.}, {Hivon, E.}, {Huang, Z.}, {Jaffe, A. H.}, {Jones, W. C.}, {Karakci, A.}, {Keihänen, E.}, {Keskitalo, R.}, {Kiiveri, K.}, {Kim, J.}, {Kisner, T. S.}, {Knox, L.}, {Krachmalnicoff, N.}, {Kunz, M.}, {Kurki-Suonio, H.}, {Lagache, G.}, {Lamarre, J.-M.}, {Lasenby, A.}, {Lattanzi, M.}, {Lawrence, C. R.}, {Le Jeune, M.}, {Lemos, P.}, {Lesgourgues, J.}, {Levrier, F.}, {Lewis, A.}, {Liguori, M.}, {Lilje, P. B.}, {Lilley, M.}, {Lindholm, V.}, {López-Caniego, M.}, {Lubin, P. M.}, {Ma, Y.-Z.}, {Macías-Pérez, J. F.}, {Maggio, G.}, {Maino, D.}, {Mandolesi, N.}, {Mangilli, A.}, {Marcos-Caballero, A.}, {Maris, M.}, {Martin, P. G.}, {Martinelli, M.}, {Martínez-González, E.}, {Matarrese, S.}, {Mauri, N.}, {McEwen, J. D.}, {Meinhold, P. R.}, {Melchiorri, A.}, {Mennella, A.},
  {Migliaccio, M.}, {Millea, M.}, {Mitra, S.}, {Miville-Deschênes, M.-A.}, {Molinari, D.}, {Montier, L.}, {Morgante, G.}, {Moss, A.}, {Natoli, P.}, {Nørgaard-Nielsen, H. U.}, {Pagano, L.}, {Paoletti, D.}, {Partridge, B.}, {Patanchon, G.}, {Peiris, H. V.}, {Perrotta, F.}, {Pettorino, V.}, {Piacentini, F.}, {Polastri, L.}, {Polenta, G.}, {Puget, J.-L.}, {Rachen, J. P.}, {Reinecke, M.}, {Remazeilles, M.}, {Renzi, A.}, {Rocha, G.}, {Rosset, C.}, {Roudier, G.}, {Rubiño-Martín, J. A.}, {Ruiz-Granados, B.}, {Salvati, L.}, {Sandri, M.}, {Savelainen, M.}, {Scott, D.}, {Shellard, E. P. S.}, {Sirignano, C.}, {Sirri, G.}, {Spencer, L. D.}, {Sunyaev, R.}, {Suur-Uski, A.-S.}, {Tauber, J. A.}, {Tavagnacco, D.}, {Tenti, M.}, {Toffolatti, L.}, {Tomasi, M.}, {Trombetti, T.}, {Valenziano, L.}, {Valiviita, J.}, {Van Tent, B.}, {Vibert, L.}, {Vielva, P.}, {Villa, F.}, {Vittorio, N.}, {Wandelt, B. D.}, {Wehus, I. K.}, {White, M.}, {White, S. D. M.}, {Zacchei, A.}, \& {Zonca, A.}}]{PlanckCollaboration2020Astron.Astrophys.}
Aghanim, N., {Akrami, Y.}, {Ashdown, M.}, {et~al.} 2020, Astron. Astrophys., 641, A6, \dodoi{10.1051/0004-6361/201833910}

\bibitem[{Arnould {et~al.}(2007)Arnould, Goriely, \& Takahashi}]{Arnould2007Phys.Rep.}
Arnould, M., Goriely, S., \& Takahashi, K. 2007, Phys. Rep., 450, 97 , \dodoi{https://doi.org/10.1016/j.physrep.2007.06.002}

\bibitem[{Beers {et~al.}(2005)Beers, Barklem, Christlieb, \& Hill}]{Beers2005Nucl.Phys.A}
Beers, T., Barklem, P., Christlieb, N., \& Hill, V. 2005, Nucl. Phys. A, 758, 595 , \dodoi{https://doi.org/10.1016/j.nuclphysa.2005.05.108}

\bibitem[{Bennett {et~al.}(2013)Bennett, Larson, Weiland, Jarosik, Hinshaw, Odegard, Smith, Hill, Gold, Halpern, Komatsu, Nolta, Page, Spergel, Wollack, Dunkley, Kogut, Limon, Meyer, Tucker, \& Wright}]{Bennett2013Astrophys.J.Suppl.Ser.}
Bennett, C.~L., Larson, D., Weiland, J.~L., {et~al.} 2013, Astrophys. J. Suppl. Ser., 208, 20, \dodoi{10.1088/0067-0049/208/2/20}

\bibitem[{Butcher(1987)}]{Butcher1987thorium}
Butcher, H. 1987, Nature, 328, 127, \dodoi{10.1038/328127a0}

\bibitem[{Caldwell \& Kamionkowski(2009)}]{Caldwell2009ARNPS}
Caldwell, R.~R., \& Kamionkowski, M. 2009, Annu. Rev. Nucl. Part. Sci., 59, 397

\bibitem[{Chen(2019)}]{Chen2019Nat.Astron.}
Chen, H.-Y. 2019, Nat. Astron., 3, 384, \dodoi{10.1038/s41550-019-0776-1}

\bibitem[{Chen {et~al.}(2021)Chen, Shi, Beers, Yan, Gao, Li, Li, \& Zhao}]{Chen2021Res.Astron.Astrophys.}
Chen, T.-Y., Shi, J.-R., Beers, T.~C., {et~al.} 2021, Res. Astron. Astrophys., 21, 036, \dodoi{10.1088/1674-4527/21/2/36}

\bibitem[{{Christlieb, N.} {et~al.}(2004){Christlieb, N.}, {Beers, T. C.}, {Barklem, P. S.}, {Bessell, M.}, {Hill, V.}, {Holmberg, J.}, {Korn, A. J.}, {Marsteller, B.}, {Mashonkina, L.}, {Qian, Y.-Z.}, {Rossi, S.}, {Wasserburg, G. J.}, {Zickgraf, F.-J.}, {Kratz, K.-L.}, {Nordström, B.}, {Pfeiffer, B.}, {Rhee, J.}, \& {Ryan, S. G.}}]{Christlieb2004Astron.Astrophys.}
{Christlieb, N.}, {Beers, T. C.}, {Barklem, P. S.}, {et~al.} 2004, Astron. Astrophys., 428, 1027, \dodoi{10.1051/0004-6361:20041536}

\bibitem[{Cowan {et~al.}(1999)Cowan, Pfeiffer, Kratz, Thielemann, Sneden, Burles, Tytler, \& Beers}]{Cowan1999Astrophys.J.}
Cowan, J.~J., Pfeiffer, B., Kratz, K.-L., {et~al.} 1999, Astrophys. J., 521, 194, \dodoi{10.1086/307512}

\bibitem[{Cowan {et~al.}(2021)Cowan, Sneden, Lawler, Aprahamian, Wiescher, Langanke, Mart\'{\i}nez-Pinedo, \& Thielemann}]{Cowan2021Rev.Mod.Phys.}
Cowan, J.~J., Sneden, C., Lawler, J.~E., {et~al.} 2021, Rev. Mod. Phys., 93, 015002, \dodoi{10.1103/RevModPhys.93.015002}

\bibitem[{Cowan {et~al.}(2002)Cowan, Sneden, Burles, Ivans, Beers, Truran, Lawler, Primas, Fuller, Pfeiffer, \& Kratz}]{Cowan2002Astrophys.J.}
Cowan, J.~J., Sneden, C., Burles, S., {et~al.} 2002, Astrophys. J., 572, 861, \dodoi{10.1086/340347}

\bibitem[{Cyburt {et~al.}(2010)Cyburt, Amthor, Ferguson, Meisel, Smith, Warren, Heger, Hoffman, Rauscher, Sakharuk, Schatz, Thielemann, \& Wiescher}]{Cyburt2010Astrophys.J.Suppl.Ser.}
Cyburt, R.~H., Amthor, A.~M., Ferguson, R., {et~al.} 2010, Astrophys. J. Suppl. Ser., 189, 240, \dodoi{10.1088/0067-0049/189/1/240}

\bibitem[{Duflo \& Zuker(1995)}]{Duflo1995Phys.Rev.C}
Duflo, J., \& Zuker, A. 1995, Phys. Rev. C, 52, R23, \dodoi{10.1103/PhysRevC.52.R23}

\bibitem[{Famiano {et~al.}(2020)Famiano, Balantekin, Kajino, Kusakabe, Mori, \& Luo}]{Famiano2020Astrophys.J.}
Famiano, M., Balantekin, A.~B., Kajino, T., {et~al.} 2020, Astrophys. J., 898, 163, \dodoi{10.3847/1538-4357/aba04d}

\bibitem[{Farouqi {et~al.}(2010)Farouqi, Kratz, Pfeiffer, Rauscher, Thielemann, \& Truran}]{Farouqi2010Astrophys.J.}
Farouqi, K., Kratz, K.-L., Pfeiffer, B., {et~al.} 2010, Astrophys. J., 712, 1359, \dodoi{10.1088/0004-637x/712/2/1359}

\bibitem[{{Fischer, T.} {et~al.}(2010){Fischer, T.}, {Whitehouse, S. C.}, {Mezzacappa, A.}, {Thielemann, F.-K.}, \& {Liebend\"orfer, M.}}]{Fischer2010Astron.Astrophys.}
{Fischer, T.}, {Whitehouse, S. C.}, {Mezzacappa, A.}, {Thielemann, F.-K.}, \& {Liebend\"orfer, M.} 2010, Astron. Astrophys., 517, A80, \dodoi{10.1051/0004-6361/200913106}

\bibitem[{{Fowler} \& {Hoyle}(1960)}]{Fowler1960Ann.Phys.}
{Fowler}, W.~A., \& {Hoyle}, F. 1960, Ann. Phys., 10, 280, \dodoi{10.1016/0003-4916(60)90025-7}

\bibitem[{Frebel {et~al.}(2007)Frebel, Christlieb, Norris, Thom, Beers, \& Rhee}]{Frebel2007Astrophys.J.}
Frebel, A., Christlieb, N., Norris, J.~E., {et~al.} 2007, Astrophys. J., 660, L117, \dodoi{10.1086/518122}

\bibitem[{Gamow(1946)}]{Gamow1946Phys.Rev.}
Gamow, G. 1946, Phys. Rev., 70, 572, \dodoi{10.1103/PhysRev.70.572.2}

\bibitem[{{Goriely, S.} \& {Arnould, M.}(2001)}]{Goriely2001Astron.Astrophys.}
{Goriely, S.}, \& {Arnould, M.} 2001, Astron. Astrophys., 379, 1113, \dodoi{10.1051/0004-6361:20011368}

\bibitem[{Guo {et~al.}(2024{\natexlab{a}})Guo, Cao, Chen, Chen, Cheoun, Choi, Lam, Deng, Dong, Du, Du, Duan, Fan, Gao, Geng, Ha, He, Hu, Huang, Huang, Huang, Huang, Hyung, Chan, Jiang, Kim, Kim, Lee, Lee, Li, Li, Li, Li, Lian, Liang, Liu, Lu, Liu, Meng, Meng, Mun, Niu, Niu, Pan, Peng, Qu, Papakonstantinou, Shang, Shang, Shen, Shen, Sun, Sun, Wang, Wang, Wang, Wang, Wu, Wu, Wu, Xia, Xie, Yao, Ip, Yiu, Yu, Yu, Zhang, Zhang, Zhang, Zhang, Zhang, Zhang, Zhang, Zhang, Zhang, Zhao, Zhao, Zheng, Zhou, Zhou, \& Zou}]{Guo2024ADNDT}
Guo, P., Cao, X., Chen, K., {et~al.} 2024{\natexlab{a}}, Atom. Data Nucl. Data Tables, 158, 101661, \dodoi{https://doi.org/10.1016/j.adt.2024.101661}

\bibitem[{Guo {et~al.}(2024{\natexlab{b}})Guo, Yu, Wu, Pan, \& Zhang}]{Guo2024PRC}
Guo, Y.~Y., Yu, T., Wu, X.~H., Pan, C., \& Zhang, K.~Y. 2024{\natexlab{b}}, Phys. Rev. C, 110, 064310, \dodoi{10.1103/PhysRevC.110.064310}

\bibitem[{Hansen {et~al.}(2018)Hansen, Holmbeck, Beers, Placco, Roederer, Frebel, Sakari, Simon, \& Thompson}]{Hansen2018Astrophys.J.}
Hansen, T.~T., Holmbeck, E.~M., Beers, T.~C., {et~al.} 2018, Astrophys. J., 858, 92, \dodoi{10.3847/1538-4357/aabacc}

\bibitem[{Heng {et~al.}(2014)Heng, Xu, Niu, Sun, \& Guo}]{Heng2014JPG}
Heng, T., Xu, X., Niu, Z., Sun, B., \& Guo, J. 2014, J. Phys. G-Nucl. Part. Phys., 41, 105202, \dodoi{10.1088/0954-3899/41/10/105202}

\bibitem[{Hill {et~al.}(2017)Hill, Christlieb, Beers, Barklem, Kratz, Nordstr{\"o}m, Pfeiffer, \& Farouqi}]{Hill2017Astron.Astrophys.}
Hill, V., Christlieb, N., Beers, T.~C., {et~al.} 2017, Astron. Astrophys., 607, A91, \dodoi{10.1051/0004-6361/201629092}

\bibitem[{Hill {et~al.}(2002)Hill, Plez, Cayrel, Beers, Nordstr{\"o}m, Andersen, Spite, Spite, Barbuy, Bonifacio, Depagne, Francois, \& Primas}]{Hill2002Astron.Astrophys.}
Hill, V., Plez, B., Cayrel, R., {et~al.} 2002, Astron. Astrophys., 387, 560, \dodoi{10.1051/0004-6361:20020434}

\bibitem[{Holmbeck {et~al.}(2018)Holmbeck, Beers, Roederer, Placco, Hansen, Sakari, Sneden, Liu, Lee, Cowan, \& Frebel}]{Holmbeck2018Astrophys.J.}
Holmbeck, E.~M., Beers, T.~C., Roederer, I.~U., {et~al.} 2018, Astrophys. J., 859, L24, \dodoi{10.3847/2041-8213/aac722}

\bibitem[{Jee {et~al.}(2019)Jee, Suyu, Komatsu, Fassnacht, Hilbert, \& Koopmans}]{Jee2019Science}
Jee, I., Suyu, S.~H., Komatsu, E., {et~al.} 2019, Science, 365, 1134, \dodoi{10.1126/science.aat7371}

\bibitem[{Jiang {et~al.}(2024)Jiang, Wu, Zhao, \& Meng}]{Jiang2024PLB}
Jiang, X., Wu, X., Zhao, P., \& Meng, J. 2024, Phys. Lett. B, 849, 138448, \dodoi{https://doi.org/10.1016/j.physletb.2024.138448}

\bibitem[{Jiang {et~al.}(2021)Jiang, Wu, \& Zhao}]{Jiang2021Astrophys.J.}
Jiang, X.~F., Wu, X.~H., \& Zhao, P.~W. 2021, Astrophys. J., 915, 29, \dodoi{10.3847/1538-4357/ac042f}

\bibitem[{Kajino {et~al.}(2019)Kajino, Aoki, Balantekin, Diehl, Famiano, \& Mathews}]{Kajino2019Prog.Part.Nucl.Phys.}
Kajino, T., Aoki, W., Balantekin, A., {et~al.} 2019, Prog. Part. Nucl. Phys., 107, 109 , \dodoi{https://doi.org/10.1016/j.ppnp.2019.02.008}

\bibitem[{Kobayashi {et~al.}(2020)Kobayashi, Karakas, \& Lugaro}]{Kobayashi2020Astrophys.J.}
Kobayashi, C., Karakas, A.~I., \& Lugaro, M. 2020, Astrophys. J., 900, 179, \dodoi{10.3847/1538-4357/abae65}

\bibitem[{Koning {et~al.}(2007)Koning, Hilaire, \& Duijvestijn}]{Koning2007}
Koning, A.~J., Hilaire, S., \& Duijvestijn, M. 2007, in ND Conference, EDP Sciences, 211--214, \dodoi{10.1051/ndata:07767}

\bibitem[{Kratz {et~al.}(2007)Kratz, Farouqi, Pfeiffer, Truran, Sneden, \& Cowan}]{Kratz2007Astrophys.J.}
Kratz, K.-L., Farouqi, K., Pfeiffer, B., {et~al.} 2007, Astrophys. J., 662, 39, \dodoi{10.1086/517495}

\bibitem[{Li {et~al.}(2019)Li, Niu, \& Sun}]{Li2019Sci.ChinaPhys.Mech.Astron.}
Li, Z., Niu, Z., \& Sun, B. 2019, Sci. China-Phys. Mech. Astron., 62, 982011, \dodoi{10.1007/s11433-018-9355-y}

\bibitem[{Lodders(2003)}]{Lodders2003Astrophys.J.}
Lodders, K. 2003, Astrophys. J., 591, 1220, \dodoi{10.1086/375492}

\bibitem[{Meyer(2013)}]{Meyer2013Proc.ofScienceNICXII}
Meyer, B.~S. 2013, Proc. of Science (NIC XII), 146, 096, \dodoi{10.22323/1.146.0096}

\bibitem[{M\"oller {et~al.}(2003)M\"oller, Pfeiffer, \& Kratz}]{Moeller2003Phys.Rev.C}
M\"oller, P., Pfeiffer, B., \& Kratz, K.-L. 2003, Phys. Rev. C, 67, 055802, \dodoi{10.1103/PhysRevC.67.055802}

\bibitem[{Mumpower {et~al.}(2016)Mumpower, Surman, McLaughlin, \& Aprahamian}]{Mumpower2016Prog.Part.Nucl.Phys.}
Mumpower, M., Surman, R., McLaughlin, G., \& Aprahamian, A. 2016, Prog. Part. Nucl. Phys., 86, 86 , \dodoi{https://doi.org/10.1016/j.ppnp.2015.09.001}

\bibitem[{Nakamura {et~al.}(2015)Nakamura, Kajino, Mathews, Sato, \& Harikae}]{Nakamura2015Astron.Astrophys.}
Nakamura, K., Kajino, T., Mathews, G.~J., Sato, S., \& Harikae, S. 2015, Astron. Astrophys., 582, A34, \dodoi{10.1051/0004-6361/201526110}

\bibitem[{Nishimura {et~al.}(2012)Nishimura, Kajino, Mathews, Nishimura, \& Suzuki}]{Nishimura2012Phys.Rev.C}
Nishimura, N., Kajino, T., Mathews, G.~J., Nishimura, S., \& Suzuki, T. 2012, Phys. Rev. C, 85, 048801, \dodoi{10.1103/PhysRevC.85.048801}

\bibitem[{Nishimura {et~al.}(2015)Nishimura, Takiwaki, \& Thielemann}]{Nishimura2015Astrophys.J.}
Nishimura, N., Takiwaki, T., \& Thielemann, F.-K. 2015, Astrophys. J., 810, 109, \dodoi{10.1088/0004-637x/810/2/109}

\bibitem[{Niu {et~al.}(2009)Niu, Sun, \& Meng}]{Niu2009Phys.Rev.C}
Niu, Z., Sun, B., \& Meng, J. 2009, Phys. Rev. C, 80, 065806, \dodoi{10.1103/PhysRevC.80.065806}

\bibitem[{Niu \& Liang(2018)}]{Niu2018Phys.Lett.B}
Niu, Z.~M., \& Liang, H.~Z. 2018, Phys. Lett. B, 778, 48 , \dodoi{10.1016/j.physletb.2018.01.002}

\bibitem[{Niu \& Liang(2022)}]{Niu2022Phys.Rev.C}
---. 2022, Phys. Rev. C, 106, L021303, \dodoi{10.1103/PhysRevC.106.L021303}

\bibitem[{Otsuki {et~al.}(2003)Otsuki, Mathews, \& Kajino}]{Otsuki2003NewAstron.}
Otsuki, K., Mathews, G.~J., \& Kajino, T. 2003, New Astron., 8, 767 , \dodoi{https://doi.org/10.1016/S1384-1076(03)00065-4}

\bibitem[{Pan {et~al.}(2022)Pan, Cheoun, Choi, Dong, Du, Fan, Gao, Geng, Ha, He, Huang, Huang, Kim, Kim, Lee, Lee, Li, Liu, Ma, Meng, Mun, Niu, Papakonstantinou, Shang, Shen, Shen, Sun, Sun, Wu, Wu, Xia, Yan, Yiu, Zhang, Zhang, Zhang, Zhang, Zhao, Zheng, \& Zhou}]{Pan2022Phys.Rev.C}
Pan, C., Cheoun, M.-K., Choi, Y.-B., {et~al.} 2022, Phys. Rev. C, 106, 014316, \dodoi{10.1103/PhysRevC.106.014316}

\bibitem[{Peebles \& Ratra(2003)}]{Peebles2003Rev.Mod.Phys.}
Peebles, P. J.~E., \& Ratra, B. 2003, Rev. Mod. Phys., 75, 559, \dodoi{10.1103/RevModPhys.75.559}

\bibitem[{{Pian} {et~al.}(2017){Pian}, {D'Avanzo}, {Benetti}, {Branchesi}, {Brocato}, {Campana}, {Cappellaro}, {Covino}, {D'Elia}, {Fynbo}, {Getman}, {Ghirland a}, {Ghisellini}, {Grado}, {Greco}, {Hjorth}, {Kouveliotou}, {Levan}, {Limatola}, {Malesani}, {Mazzali}, {Melandri}, {M{\o}ller}, {Nicastro}, {Palazzi}, {Piranomonte}, {Rossi}, {Salafia}, {Selsing}, {Stratta}, {Tanaka}, {Tanvir}, {Tomasella}, {Watson}, {Yang}, {Amati}, {Antonelli}, {Ascenzi}, {Bernardini}, {Bo{\"e}r}, {Bufano}, {Bulgarelli}, {Capaccioli}, {Casella}, {Castro-Tirado}, {Chassande-Mottin}, {Ciolfi}, {Copperwheat}, {Dadina}, {De Cesare}, {di Paola}, {Fan}, {Gendre}, {Giuffrida}, {Giunta}, {Hunt}, {Israel}, {Jin}, {Kasliwal}, {Klose}, {Lisi}, {Longo}, {Maiorano}, {Mapelli}, {Masetti}, {Nava}, {Patricelli}, {Perley}, {Pescalli}, {Piran}, {Possenti}, {Pulone}, {Razzano}, {Salvaterra}, {Schipani}, {Spera}, {Stamerra}, {Stella}, {Tagliaferri}, {Testa}, {Troja}, {Turatto}, {Vergani}, \& {Vergani}}]{Pian2017Nature}
{Pian}, E., {D'Avanzo}, P., {Benetti}, S., {et~al.} 2017, Nature, 551, 67, \dodoi{10.1038/nature24298}

\bibitem[{Piatti {et~al.}(2017)Piatti, Aparicio, \& Hidalgo}]{Piatti2017Mon.Not.Roy.Astron.Soc.}
Piatti, A.~E., Aparicio, A., \& Hidalgo, S.~L. 2017, Mon. Not. Roy. Astron. Soc., 469, 1175, \dodoi{10.1093/Mon. Not. Roy. Astron. Soc./stx1002}

\bibitem[{Placco {et~al.}(2017)Placco, Holmbeck, Frebel, Beers, Surman, Ji, Ezzeddine, Points, Kaleida, Hansen, Sakari, \& Casey}]{Placco2017Astrophys.J.}
Placco, V.~M., Holmbeck, E.~M., Frebel, A., {et~al.} 2017, Astrophys. J., 844, 18, \dodoi{10.3847/1538-4357/aa78ef}

\bibitem[{Schatz {et~al.}(2002)Schatz, Toenjes, Pfeiffer, Beers, Cowan, Hill, \& Kratz}]{Schatz2002Astrophys.J.}
Schatz, H., Toenjes, R., Pfeiffer, B., {et~al.} 2002, Astrophys. J., 579, 626, \dodoi{10.1086/342939}

\bibitem[{Siegel {et~al.}(2019)Siegel, Barnes, \& Metzger}]{Siegel2019Nature}
Siegel, D.~M., Barnes, J., \& Metzger, B.~D. 2019, Nature, 569, 241, \dodoi{10.1038/s41586-019-1136-0}

\bibitem[{Sneden {et~al.}(2009)Sneden, Lawler, Cowan, Ivans, \& Hartog}]{Sneden2009Astrophys.J.Suppl.Ser.}
Sneden, C., Lawler, J.~E., Cowan, J.~J., Ivans, I.~I., \& Hartog, E. A.~D. 2009, Astrophys. J. Suppl. Ser., 182, 80, \dodoi{10.1088/0067-0049/182/1/80}

\bibitem[{Thielemann {et~al.}(1993)Thielemann, Bitouzet, Kratz, M{\"o}ller, Cowan, \& Truran}]{Thielemann1993PR}
Thielemann, F.-K., Bitouzet, J.-P., Kratz, K.-L., {et~al.} 1993, Phys. Rep., 227, 269, \dodoi{10.1016/0370-1573(93)90072-L}

\bibitem[{Thielemann {et~al.}(2011)Thielemann, Arcones, Käppeli, Liebendörfer, Rauscher, Winteler, Fröhlich, Dillmann, Fischer, Martinez-Pinedo, Langanke, Farouqi, Kratz, Panov, \& Korneev}]{Thielemann2011Prog.Part.Nucl.Phys.}
Thielemann, F.-K., Arcones, A., Käppeli, R., {et~al.} 2011, Prog. Part. Nucl. Phys., 66, 346 , \dodoi{https://doi.org/10.1016/j.ppnp.2011.01.032}

\bibitem[{Tian {et~al.}(2025)Tian, Li, Fang, \& Niu}]{Tian2025CPC}
Tian, L., Li, W.-F., Fang, J.-Y., \& Niu, Z.-M. 2025, Chin. Phys. C, 49, 044110, \dodoi{10.1088/1674-1137/ad9303}

\bibitem[{Wang {et~al.}(2012)Wang, Audi, Wapstra, Kondev, MacCormick, Xu, \& Pfeiffer}]{Wang2012Chin.Phys.C}
Wang, M., Audi, G., Wapstra, A., {et~al.} 2012, Chin. Phys. C, 36, 1603, \dodoi{10.1088/1674-1137/36/12/003}

\bibitem[{Wang {et~al.}(2014)Wang, Liu, Wu, \& Meng}]{Wang2014Phys.Lett.B}
Wang, N., Liu, M., Wu, X., \& Meng, J. 2014, Phys. Lett. B, 734, 215 , \dodoi{https://doi.org/10.1016/j.physletb.2014.05.049}

\bibitem[{Watson {et~al.}(2019)Watson, Hansen, Selsing, Koch, Malesani, Andersen, Fynbo, Arcones, Bauswein, Covino, {et~al.}}]{Watson2019Nature}
Watson, D., Hansen, C.~J., Selsing, J., {et~al.} 2019, Nature, 574, 497, \dodoi{doi.org/10.1038/s41586-019-1676-3}

\bibitem[{{Woosley} {et~al.}(1994){Woosley}, {Wilson}, {Mathews}, {Hoffman}, \& {Meyer}}]{Woosley1994Astrophys.J.}
{Woosley}, S.~E., {Wilson}, J.~R., {Mathews}, G.~J., {Hoffman}, R.~D., \& {Meyer}, B.~S. 1994, Astrophys. J., 433, 229, \dodoi{10.1086/174638}

\bibitem[{Wu {et~al.}(2021)Wu, Guo, \& Zhao}]{Wu2021Phys.Lett.B}
Wu, X.~H., Guo, L.~H., \& Zhao, P.~W. 2021, Phys. Lett. B, 819, 136387, \dodoi{https://doi.org/10.1016/j.physletb.2021.136387}

\bibitem[{Wu {et~al.}(2022{\natexlab{a}})Wu, Lu, \& Zhao}]{Wu2022Phys.Lett.B137394}
Wu, X.~H., Lu, Y.~Y., \& Zhao, P.~W. 2022{\natexlab{a}}, Phys. Lett. B, 834, 137394, \dodoi{https://doi.org/10.1016/j.physletb.2022.137394}

\bibitem[{Wu \& Meng(2023)}]{Wu2023Sci.Bull.}
Wu, X.-H., \& Meng, J. 2023, Sci. Bull., 68, 539, \dodoi{https://doi.org/10.1016/j.scib.2023.03.004}

\bibitem[{Wu \& Pan(2024)}]{Wu2024PRC_AKRR}
Wu, X.~H., \& Pan, C. 2024, Phys. Rev. C, 110, 034322, \dodoi{10.1103/PhysRevC.110.034322}

\bibitem[{Wu {et~al.}(2024)Wu, Pan, Zhang, \& Hu}]{Wu2024Phys.Rev.C}
Wu, X.~H., Pan, C., Zhang, K.~Y., \& Hu, J. 2024, Phys. Rev. C, 109, 024310, \dodoi{10.1103/PhysRevC.109.024310}

\bibitem[{Wu \& Zhao(2020)}]{Wu2020Phys.Rev.C051301}
Wu, X.~H., \& Zhao, P.~W. 2020, Phys. Rev. C, 101, 051301 (R), \dodoi{10.1103/PhysRevC.101.051301}

\bibitem[{Wu {et~al.}(2022{\natexlab{b}})Wu, Zhao, Zhang, \& Meng}]{Wu2022Astrophys.J.152}
Wu, X.~H., Zhao, P.~W., Zhang, S.~Q., \& Meng, J. 2022{\natexlab{b}}, Astrophys. J., 941, 152, \dodoi{10.3847/1538-4357/aca526}

\bibitem[{Zhang {et~al.}(2022)Zhang, Cheoun, Choi, Chong, Dong, Dong, Du, Geng, Ha, He, Heo, Ho, In, Kim, Kim, Lee, Lee, Li, Li, Luo, Meng, Mun, Niu, Pan, Papakonstantinou, Shang, Shen, Shen, Sun, Sun, Tam, Thaivayongnou, Wang, Wang, Wong, Wu, Wu, Xia, Yan, Yeung, Yiu, Zhang, Zhang, Zhang, Zhao, \& Zhou}]{Zhang2022Atom.DataNucl.DataTables}
Zhang, K., Cheoun, M.-K., Choi, Y.-B., {et~al.} 2022, Atom. Data Nucl. Data Tables, 144, 101488, \dodoi{https://doi.org/10.1016/j.adt.2022.101488}

\bibitem[{Zhang {et~al.}(2025)Zhang, Pan, Wu, Qu, Lu, \& Sun}]{Zhang2025AAPPSBulletin}
Zhang, K.~Y., Pan, C., Wu, X.~H., {et~al.} 2025, AAPPS Bulletin, 35, 13, \dodoi{10.1007/s43673-025-00153-x}

\bibitem[{Zhao \& Zhang(2019)}]{Zhao2019Astrophys.J.}
Zhao, B., \& Zhang, S.~Q. 2019, Astrophys. J., 874, 5, \dodoi{10.3847/1538-4357/ab0702}

\bibitem[{Zhou {et~al.}(2017)Zhou, Li, Wang, Chen, Guo, Su, Li, Yan, Li, Han, Shen, Gan, Zeng, Lian, \& Liu}]{Zhou2017Sci.ChinaPhys.Mech.Astron.}
Zhou, Y., Li, Z., Wang, Y., {et~al.} 2017, Sci. China-Phys. Mech. Astron., 60, 082012, \dodoi{10.1007/s11433-017-9045-0}

\end{thebibliography}
\bibliographystyle{aasjournal}

\end{CJK*}
\end{document}